# Simulated epidemics in an empirical spatiotemporal network of 50,185 sexual contacts

Luis E. C. Rocha[1], Fredrik Liljeros[2] and Petter Holme[1]

[1]IceLab, Department of Physics, Umeå University, 90187 Umeå, Sweden

[2]Department of Sociology, Stockholm University, 10691 Stockholm, Sweden

June 9, 2010

We study implications of the dynamical and spatial contact structure between Brazilian escorts and sex-buyers for the spreading of sexually transmitted infections (STI). Despite a highly skewed degree distribution diseases spreading in this contact structure have rather well-defined epidemic thresholds. Temporal effects create a broad distribution of outbreak sizes even if the transmission probability is taken to the hypothetical value of 100%. Temporal correlations speed up outbreaks, especially in the early phase, compared to randomized contact structures. The time-ordering and the network topology, on the other hand, slow down the epidemics. Studying compartmental models we show that the contact structure can probably not support the spread of HIV, not even if individuals were sexually active during the acute infection. We investigate hypothetical means of containing an outbreak and find that travel restrictions are about as efficient as removal of the vertices of highest degree. In general, the type of commercial sex we study seems not like a major factor in STI epidemics.

**1. INTRODUCTION**

Spatiotemporal structural variations in sexual contact patterns are thought to influence the spread of sexually transmitted infections. Since epidemics can be a society-wide phenomenon, and sexual contact patterns can have structure at all scales, we need population-level sexual network data to understand STI epidemics. Unfortunately it is hard to collect sexual contact data on that large a scale. Instead people have focused on small-scale studies by interviews (Bearman *et al.* 2004; Canright *et al.* 2009) or contact tracing (Potterat *et al.* 2002; Poulin *et al.* 2000; Wylie and Jolly 2001; De *et al.* 2004; Rothenberg *et al.* 1998); or they have studied larger sample sets by random sampling surveys (Liljeros *et al.* 2003; Latora *et al.* 2006; de Blasio *et al.* 2007; Ewald 2006). Small surveys and contact tracing are in danger of missing large-scale structures (Ewald 2006) and emergent phenomena. Large-scale surveys, on the other hand, mainly collect the number of partners, but not the connections between them. An alternative way to gather information about the sexual contact patterns that covers a large number of people and explicitly maps their connections is to use Internet data. In our study we use a dataset of claimed sexual contacts between Brazilian escorts (high-end prostitutes) and sex-buyers (Rocha *et al.* 2010). Contact patterns of commercial sex can of course not be generalized to a whole population but instead contain information about the time and location of the contacts that spans six years and over 16,748 individuals.

Sexual contact patterns have temporal correlations both at an individual and population level (Rocha *et al.* 2010). Just like the network structure, temporal structures could influence epidemics in several ways. For example, consider three agents, A, B and C where B and C are in contact first and later A and B. Considering the temporal order of the contacts disease cannot spread from A to C via B, but in a static network representation this information is lost so C will appear reachable from A via B (Riolo *et al.* 2001; Kempe *et al.* 2002; Moody 2002; Holme 2005; Volz & Meyers 2007). A conspicuous temporal structure in human communication that we also observe in our data is bursty behaviour where people are very active for a limited period at a time (Eckmann *et al.* 2004). Another example of a temporal structure is the long term behavioural change where new individuals enter the system and others leave. To investigate such temporal effects empirically requires time stamps on the contacts; and again Internet data sets such as ours, as opposed to most above-mentioned data, have such information.

Extending epidemiological models to include space is a common step to include structure beyond the well-mixed assumption (Elliot *et al.* 2000; Lawson 2001). Geography has several imprints on the contact structure and thus disease spreading – it makes contact network becomes larger than a random graph in terms of graph distances; it create network clusters corresponding to densely populated areas (Rocha *et al.* 2010).

This stress the importance of network-data sets covering a large area of space. Our data set covers, we believe, the escort business of Brazil rather completely (Rocha *et al.* 2010).

In this work we address the question how the contact structure in the contact data of Rocha *et al.* (2010) affect epidemic spreading in general, and with HIV and Human papillomavirus as special examples. For most of our study we look at spreading processes confined to our contact data. Because of the lack of similar data on other types of sexual interaction, it is hard to draw conclusions on the role of escort business on the spread of STIs in the society as a whole. Rather we investigate the role of topological, temporal and geographic structure on transmission pathways within this type of commercial sex. We do however make a crude estimate of the role of commercial sex in a population-wide context. Furthermore we briefly investigate some strategies to limit epidemic outbreaks.

## 2. OUR EMPIRICAL SEXUAL NETWORK

In this section, we will briefly describe the community and describe how we use it to derive a network of claimed sexual contacts (this is described in more detail in Rocha *et al.* 2010). The community is a public online forum, oriented to heterosexual males (sex-buyers), who evaluate and comment their sexual encounters with female prostitutes (sex-sellers), both using anonymous aliases. The posts in the forum are organised by the location of the encounter and by type of prostitution (defined by the price-level and mode of acquiring customers). We focus in the escorts (the most expensive form of prostitution; Edlund *et al.* 2009) section of the forum, mostly because it is more organized than the other sections. Connecting a sex-buyer posting in a forum thread to the escort (that is the topic of the thread) creates a bipartite network where the edges are tagged with the times of the posts. We take the times of the posts as estimates of the times of the sexual encounters (even though, by inspection, it is clear that the sex-buyers often post about several encounters at the same session). The posts also include information about the types of sexual activity that were performed in three categories: oral sex without condom, kiss on mouth and anal sex. The dataset covers the beginning of the community, September 2002 through October 2008. All in all there are 50,185 contacts recorded between 6,642 escorts and 10,106 sex-buyers. These contacts make up a network with a largest connected cluster covering over 97% of the individuals (despite that the network is spread out over twelve cities of Brazil).

## 3. SIMULATION OF EPIDEMICS WITHIN THE DATA

In this section, we simulate disease spreading within the recorded and try to relate the observations to the network structure. We mostly use the Susceptible Infected (SI) model, both for the sake of parsimony and also because it is relevant for diseases with long infectious stages (like HIV or Human papillomavirus, HPV). To account for short infectious stages, we also simulate the Susceptible Infected Removed (SIR) model. This section is not an attempt to cover all aspects of disease spreading on this underlying structure, but rather aiming at investigating synergetic effects of network structure and contact dynamics within this contact sample.

### 3.1. SI MODEL

In the SI model, a vertex can be in two states: susceptible or infective. Once infective, the state does not change. This can be a crude model for e.g. HIV and other infections with infective periods longer than the timescale of the contact dynamics. The probability that the infection will spread during an encounter – the transmission probability, $\rho$ – is the only independent parameter of the model. For the simulations we present here, we run through the contact list from older to newer contacts. Since we only have information about the day on which the encounter was reported, we take averages over 100 random orderings of contacts reported on the same day. Strictly speaking, the order does not have to be the same as the actual sexual encounters, but as the sampling time gets longer this should matter less, and it should be seen in context of the otherwise large approximations of compartmental models such as the SI model.

Since the community is evolving, no matter how we define the static network structure, it will change as time goes on. To minimize finite-size effects, we skip the initial 1000 days in the data which correspond to a transient period with fewer users and sparse encounters (it is later shown that 1000 days is an adequate choice). In the initial configuration, we let all vertices be susceptible except for one randomly selected sex-seller acting as source of the infection and run the simulations within a window of 800 days. Only individuals having encounters within the interval from $T = 1000$ days to $T = 1200$ days can be selected. In addition to the averages over different contact orders during the same day, we average 50 times for each initial condition (we expect a larger dependence on the initial conditions for sparser networks, or lower transmission rates). Before taking the averages, we normalize the number of infected vertices by the total number in the sample. The average number of vertices (the sample population) within the window of size 800 days is $10{,}526 \pm 145$ and the number of edges (or posts) is $32{,}685 \pm 378$.

A straightforward way of investigating the effects of time ordering is to remove different types of temporal correlations by randomizing the contact sequence in different ways. In figure 1 we investigate effects of the time ordering of contacts by the SI model and $\rho = 1$ comparing the original data with three different randomizations. In our figure 1*a* we randomly swap the time-stamps of the contacts between themselves, keeping everything else like in the original data (so that the pairs of sex-sellers and -buyers, the number of contacts involving such a pair, and the set of the 32,685 time stamps all are conserved, but the time an encounter happen is unrelated to the original data). We see that an infection spreads much slower in the randomized than the original data, reaching less than 50% of the individuals compared to over 60% for the original data. Thus, correlations in the order the contacts occur in the empirical data speeds up disease spreading. More concretely, one such tendency is that individuals tend to be intensely active over a period of time, then quit completely- When the time stamps are randomized, this tendency will disappear and the presence of individuals in the system will, on average, be longer and the contact frequency sparser. The average time between the first and last occurrence in the data increase from 170.9 to $337.5 \pm 0.1$ days by the randomization. In addition to correlations in the temporal order of contacts, the topology of the sexual network can influence epidemics (De *et al.* 2004; Rothenberg *et al.* 1998; Potterat *et al.* 2002; Liljeros *et al.* 2003; Volz & Meyers 2007). One network structure that, specifically, is known to affect disease spreading is clustering—the density of triangles in the network. Many triangles can change an estimated exponential growth to polynomial (Szendroi 2004). In our present data, being purely heterosexual, there are no triangles but a large amount of four-cycles (four times more than in randomized networks). This was explained as a side effect of a strong geographic clustering in Rocha *et al.* (2010), where sex-sellers and -buyers form sparsely interconnected, rather dense clusters. In fig. 1b we compare the epidemics on the empirical data with that of a randomized data set where the degrees and bipartite structures are fixed, as are the time stamps of the escorts, but the pairs are randomized. The disease outbreak seems to be slowed down, at least during an initial 200 days, by the network topology as the topologically randomized data gives faster and more pervasive outbreaks in fig. 1b. The same is true if the time stamps of the sex-buyers are conserved. That the epidemics spread faster initially on the original data come from the locally high clustering within cities. Finally we randomize both the contact times and the contact pairs altogether, see fig. 1c. With both the temporal and topological information randomized, the curve is in between those of fig. 1a and fig. 1b. The fraction of infected vertices has a slow increase during the initial 300 days, but not slower than the temporally randomized data of fig. 1a. Later it increases faster and reaches, by the end of the sampling period, about 70% of the individuals (a little less than for the topology-only randomization, but larger than for the original data).

In figure 2, we investigate the effects of lowering the transmission rate from the maximal value, $\rho = 1$, used in fig. 1, to more realistic values (even though e.g. HPV has a very high transmission probability; D'Souza *et al*. 2007). We measure the average outbreak size relative to the size with a maximal transmission rate. The relative fraction of infected individuals decreases to a minimum at about 100 days. The time scale of the position of the minimum, which corresponds to the time lag of secondary infection pathways, is more pronounced for lower transmission rates. The fact that the curves are fairly constant (differs at most 25% for $\rho = 0.6$) is an indication that our results for the $\rho = 1$ case holds for other transmission rates as well, i.e. that the time-ordering effects are stronger than the fluctuations from the stochasticity of the contagion process.

In figure 3 we measure the probability distribution of outbreak sizes. We note that there is a large diversity of outbreaks even for $\rho = 1$. This, we hypothesize, is a general phenomenon – temporal constraints increase the diversity of outbreaks since it restricts the possible infection paths in the network. There is, however, a local maximum where, for $\rho = 1$, a fraction of about 0.75 of the sample population gets infected (for $\rho = 1$), which sets a characteristic outbreak size. This local maximum depends on the transmission rate and decreases for lower values (slightly superlineraly). Another observation is that the outbreak-size distribution becomes more homogeneous for lower transmission rates. The peak on the very left of the graph indicates that for most sources of disease, the disease will die out. Note that the data also contain isolated connected components that, once infected, do not spread the disease to the giant component. We also plot the outbreak distribution for the network of different types of sexual contact and note that the oral sex without condom and kiss on mouth (an important transmission pathway for e.g. Herpes Simplex Virus; Martin *et al.* 2009; and HPV; D'Souza *et al.* 2007) have roughly the same outbreak distribution (despite being about half as dense as the full contact structure).

In figure 4*a*, we investigate the average effects of varying $\rho$ and see that the average outbreak size is an approximately linear function of transmissibility. From the figure, we see that epidemic outbreak is practically absent for transmission rates lower than about 0.19—a *de facto* threshold effect. To investigate if this threshold value $\rho^*$ is an artefact of the finite-size sampling, we use different sampling windows from the complete dataset and take the crossing point between the fit (of the fraction of infected as a function of $\rho$) to a linear functional form and the line of no secondary infections as an estimate of the threshold value. Each window represents the same duration, but starts at a different time ($T_0$). We see in figure 4*b*, the apparent convergence of the threshold estimates to values about $\rho^* = 0.19 \pm 0.02$ for increasing $T_0$, which is our estimated threshold value for this contact pattern. This is above the estimated transmissibility of chronic HIV (for any type of sex) so we can conclude that our data cannot support HIV transmitted via chronically infected individuals (Royce *et al.* 1997).

**3.2. SIR MODEL**

The SI model simulates a situation where the spreading dynamic is faster than the infectious period. This is clearly unrealistic for many diseases, especially over the time scale of our data set (over six years). In such cases one adds a compartment R (removed) of individuals that cannot become re-infected but stays in R for good. If we assume the main transmission pathway of HIV is contacts during the more contagious, early, acute-infection stage lasting about four weeks (Pilcher *et al.* 2007), the HIV can be modelled by the SIR model. This hypothesis might be reasonable in the short pathogenic exposure associated with commercial sex. In this section we will test if our sexual network is capable of sustaining such an outbreak. When simulating the SIR model we follow the same protocol as the SI simulations only that vertices stay in the infective state a time δ after which they become removed.

In figure 5*a*, we plot the outbreak size as a function of the duration of the infective stage. Here we assume a maximal transmission rate ρ = 1. We proceed to identify estimated threshold values by performing fits of second-order polynomials to the fraction of non-susceptible individuals and identifying the crossing point with the no-secondary-infection line. Performing the same analysis as above for the SI model's transmissibility threshold, but now for the duration of the infective state, we find that the δ-threshold converge to δ* = 31±1 days. Since this is more than the estimated duration of an acute HIV infection and we overestimated the transmission probability, we conclude our measured contact pattern is not dense enough to support an outbreak of HIV by the acute infection pathways. In reality there are other sexual contacts, outside of the reported, that connect the individuals of our data. HIV is endemic to Brazil, so our conclusion is that other pathways (like man-to-man sex or needle sharing among injecting drug users) than Internet-mediated commercial sex is needed to explain the state of the epidemics in Brazil. Below we will discuss the effects of adding our empirical network of commercial sex to a well-mixed sexual network model.

**4. EFFECTS OF TARGETED CAMPAIGNS**

In this section, we investigate the limits of efficiency for disease prevention with targeted actions, like vaccination. For simplicity we study the SI model with ρ = 1. Figure 6*a* shows, as expected from heterogeneous network that deleting a small fraction of the vertices in order of highest degree decrease outbreak sizes considerable (e.g. 90% with a deletion of 2% of the vertices). This type of intervention could be technically possible in online communities like ours where the number of partners is possible to estimate. Another structure that could be useful, at least if one could disregard legal issues, is related to travel restrictions (Camitz & Liljeros 2006). Our dataset is, as mentioned, also highly clustered reflecting a geographic structure (Rocha *et al.* 2010). In fig. 6*b* we investigate the average largest outbreak sizes if we delete sex-buyer vertices in order of the number of connections outside of their hometown (defined as the one they have most contacts in). The response to the deletion of travellers is not as dramatic as deletion according to degree but note that the fraction of travellers is only about 5% of the total number of vertices so the two mentors are about as efficient up to 5% removed vertices.

Next, we investigate scenarios where the disease and the vertex-removal occur simultaneously. Imagine a vaccination campaign launched a time $\tau$ after the disease enters the system and the campaign manages to reach everyone active in the interval [$\tau$, $\tau + \Delta\tau$]. The variable $\tau$ parameterizes the response time of the society; $\Delta\tau$ quantifies the resources to prevent the disease spreading. In figure 6*c*, we see an approximate power-law scaling of the average outbreak size as a function of the initial time $\tau$ of the vaccination. The growth decreases with the duration of the campaign (for a short campaign, it does not matter so much when it is done as for a long campaign), but a long campaign, given $\tau$, is always better than a short. For instance, a 56 day campaign starting 20 days after the primary infection is about as efficient as a 112 day campaign starting after 100 days. In practice not everyone would be reached by such a campaign. In fig. 6*d* we investigate lower efficiencies of such campaigns. The response seems to be linear (which is expected from the linear $\rho$-dependence of outbreak size far above the threshold, fig. 4a). However, the pivot in choosing between extending the vaccination interval and vaccinating more individuals is nonlinearly dependent on the fraction of removed vertices. If, for instance, we half the fraction of vaccinated individuals (from 0.8 to 0.4) we would have to compensate that by a 4-fold increase of $\Delta\tau$ (from 28 to 112 days). A decrease of the fraction of infected from 1 to 0.6 is equivalent to increase of $\Delta\tau$ from 56 to 168 days.

## 5. AUGMENTING WELL-MIXED MODELS

Our community is not isolated from the rest of the sexual contacts. We concluded above that the observed contact structure alone is not enough to sustain a HIV epidemic either via acute or chronic infections. But if none of these contacts would have happened, how different would the disease spreading be? How much do they add to the background sexual activity? We will not go into depth in this question but simply compare the change in the basic reproduction number $R_0$ (in the simplest models $R_0$ is the number of secondary infected in completely uninfected population) if the contacts we record are added to an underlying contact pattern. In well-mixed models with identical agents, $R_0 = 1$ marks the epidemic threshold (so if $R_0 > 1$ there is a chance for epidemic outbreaks).

In order for this threshold criterion to hold for populations with a distribution of contact rates one can calculate an effective $R_0$ by multiplying by the factor

$$\Lambda = 1 + \frac{\sigma^2}{c^2}, \qquad (1)$$

where $\sigma^2$ is the variance of the contact rates and $c$ is the average contact rate (Anderson & May, 1991). We take an estimate of this factor from Liljeros *et al.* (2001) where $\sigma = 1.26(4)$ year$^{-1}$ and $c = 1.22(7)$ year$^{-1}$ (where the number in parenthesis gives the standard error in order of the last decimal) giving $\Lambda = 2.1(1)$. This number comes from a random sample of Swedes aged 18 to 74. Now, in a very crude calculation, we use the same values for our dataset covering a fraction of about 0.07% of the inhabitants in this age class in the sampled twelve Brazilian cities. We have $\sigma_+ = 16.3$ year$^{-1}$ and $c_+ = 5.0(1)$ year$^{-1}$ from our data. This means the difference $d$ between $\Lambda$ in the case with and without the claimed contacts of our data is

$$d = \frac{\sigma^2 + \nu\sigma_+^2}{(c + \nu c_+)^2} - \frac{\sigma^2}{c^2} \approx 0.1. \qquad (2)$$

The sexual contacts in our data would thus, in this sketchy calculation, contribute to about 5% of the $R_0$-correction factor and thus be rather insignificant. In figure 7, we plot $d$ as a function of $c$ and $\sigma$ for a range around the above values. If the average value above is an overestimate then $d$ could be close to $\Lambda$ meaning that the contacts we study actually, within this crude model framework, make a sizable contribution to STI epidemics.

## 6. DISCUSSION

We simulate spreading of infections on probably the largest yet recorded network of claimed sexual contacts. Our data come from a web community of sex-buyers discussing their encounters with escorts. Although the data spread out over twelve cities it is to a large extent connected so that a disease could spread from most parts of the system to most other. As any result based on a subset of a large dynamic network, we should be cautious to extrapolate our results to society at large, especially since commercial sex is driven by other mechanisms than regular sexual networks (which, as our data also suggest, gives a different contact structure). Instead of estimating thresholds and other expectation values we focus on how correlations in the empirical data would affect assumptions of well-mixed models. From studying randomizations of the SI model with 100% transmission probability we conclude that temporal correlations speed up the epidemics, especially in the early phase of superlinear growth. This can have important implications both for disease modelling, where temporal correlations in contact patterns should not be underestimated, and the design of vaccination protocols where temporal structures could potentially be used to detect important individuals to vaccinate (similar to how network structure can be exploited in vaccination protocols; Cohen *et al.* 2003). The network-topological correlations, on the other hand, slow down epidemics compared to a random contact structure. We know our network has both a high density of short cycles and community structure (reflecting that most sex-buyers buy sex in one region, presumably their hometown). Both these factors, many short clusters and distinct communities, are known to slow down diffusion in networks (Szendroi & Csanyi 2004; Eriksen *et al.* 2003; Salathé & Jones 2010).

Most of our analysis is at a general STI level, but to exemplify we consider HIV epidemics. From the SI simulations, we also learn that our empirical data cannot support an HIV outbreak in the chronic infection stage. This is, on the other hand, not expected either—a key predictor of HIV transmission is the viral load (Quinn *et al.* 2000) which is low enough during the chronic infection that it would need several contacts over a short time span between two individuals for the infection to spread (which is predictably rare in our data). Another scenario for HIV transmission is spreading during the first, acute phase where the infective subjects transfer higher viral loads. This can be modelled by an SIR model where R represents the chronic infection. Also in this case, with (the unrealistically high) 100% transmission probability per contact, HIV would probably not spread—the duration of the acute infection $\delta$ would have to be at least a month, while according to previous studies it is (in median) about two weeks. This result comes from the $\delta$-scaling of the epidemic threshold. Indeed it is a somewhat surprising that we see rather well-defined thresholds at all—graphs with power-law degree distributions (our data is close to that) are known to lack epidemic thresholds.

If one restricts the network to only certain types of encounters reported in the data (oral sex without condom and kiss on mouth), the network is still connected and in worst case scenarios capable of transmitting disease. Investigating the response of our data to vaccination, we find that deleting high-degree vertices is (as expected) an efficient way of stopping outbreaks. Deleting the most frequent travellers is about as efficient.

To epitomize our main findings: considering the contact patterns of Internet mediated prostitution, temporal correlations speed up the disease spreading and the network topology and time-order slow it down, creating a subexponential growth of the outbreak. The contact set we investigate is not enough to sustain a HIV outbreak and would, from a sketchy estimate, only affect $R_0$ of STIs by about 10%. This is in line with other studies toning down the role of prostitution as a pool of STIs (at least in more developed societies; Leclerc *et al.* 1986; D'Costa *et al.* 1985; van den Hoek *et al.* 2001; Gisselquist & Correa 2006). A future direction is to investigate how far this conclusion can be generalized—to escorts in other cultures, to other forms of commercial sex, or even non-commercial sexual contact patterns.

**Acknowledgments**

LECR and PH acknowledge financial support from the Swedish Foundation for Strategic Research and the Swedish Research Foundation. FL acknowledges Riksbankens Jubileumsfond for financial support.

**Figures**

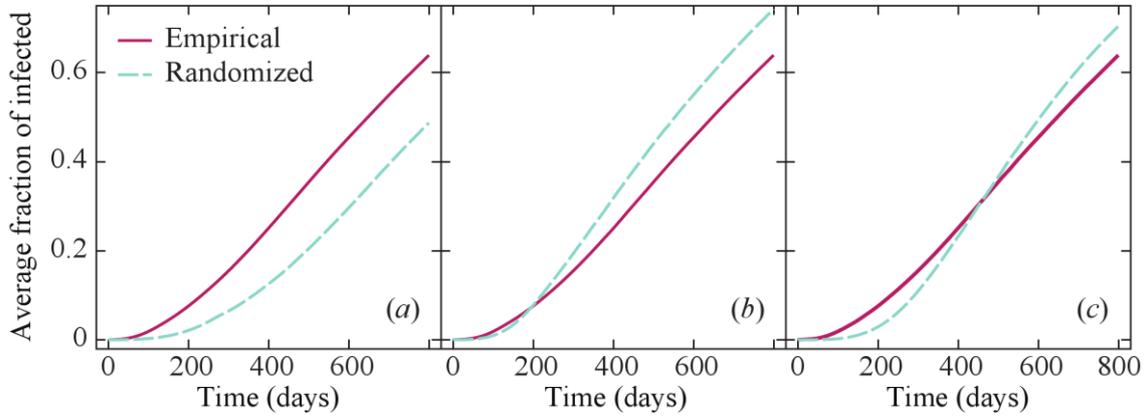

Figure 1. Effects of time ordering and temporal correlations of contacts. In (*a*)–(*c*) we plot the time evolution of the fraction of infected individuals. The curves correspond to SI epidemics in the original data (full line) and in its randomized versions: (*a*) swapping time stamps; (*b*) rewiring the edges and maintaining the sellers' time correlation; and (*c*) swapping time stamps and rewiring the edges simultaneously.

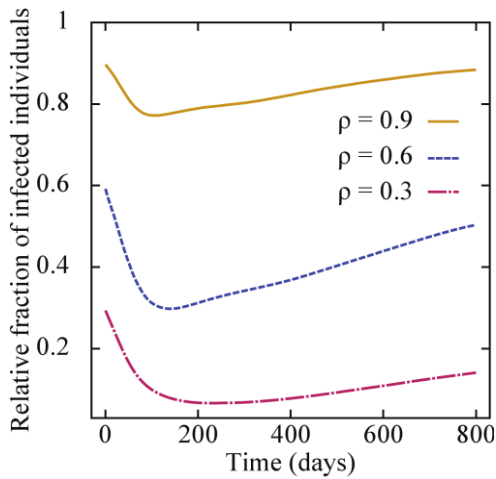

Figure 2. The relative number of infected individuals in comparison to the same value for the maximum transmission rate for different transmission rates.

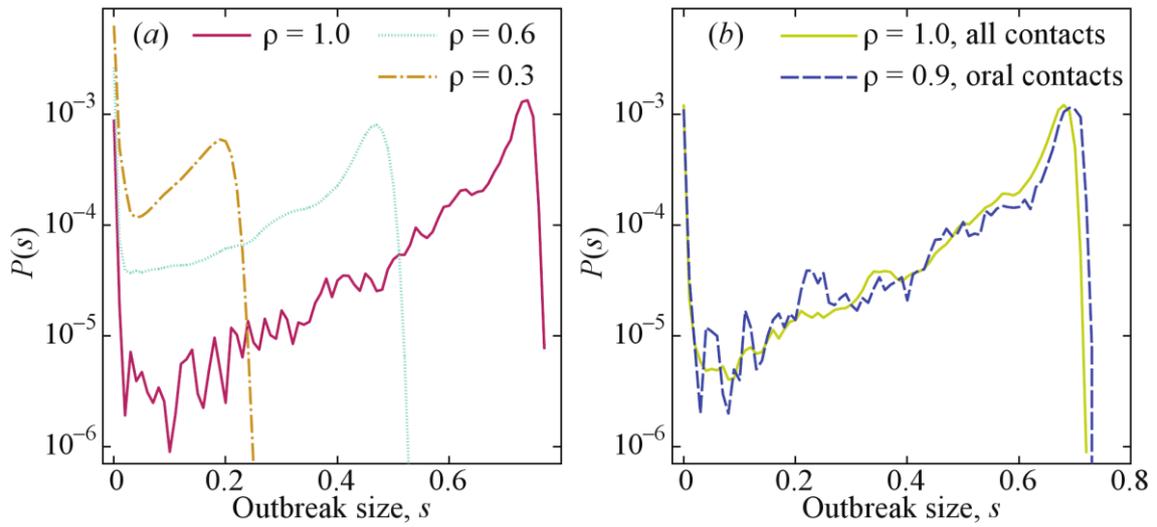

Figure 3. Comparison of different contagion pathways. We plot the probability distribution of the outbreak sizes for different transmission rates ρ for both the empirical data and the network of oral-sex and kiss-on-mouth encounters (ρ = 1).

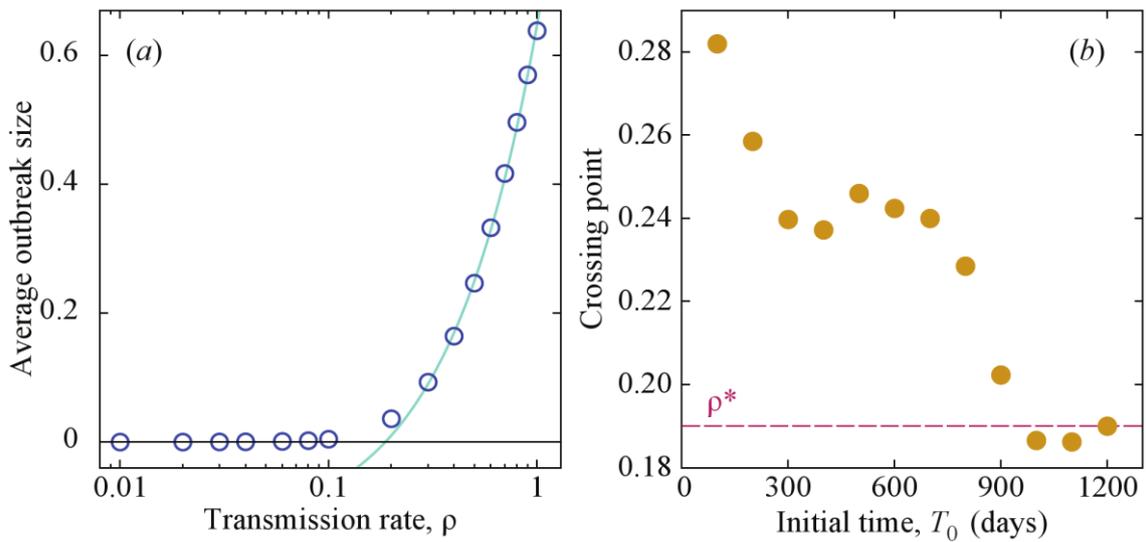

Figure 4. (*a*) displays the average outbreak size as a function of the transmission rate. The line is a linear trend least-square fitted to the data in the interval $0.3 \leq \rho \leq 1$. The abscissa is in log-scale. Panel (*b*) shows the threshold transmission rate (estimated the crossing of the linear fitting and the zero-size outbreak line) as a function of the beginning of the sampling window.

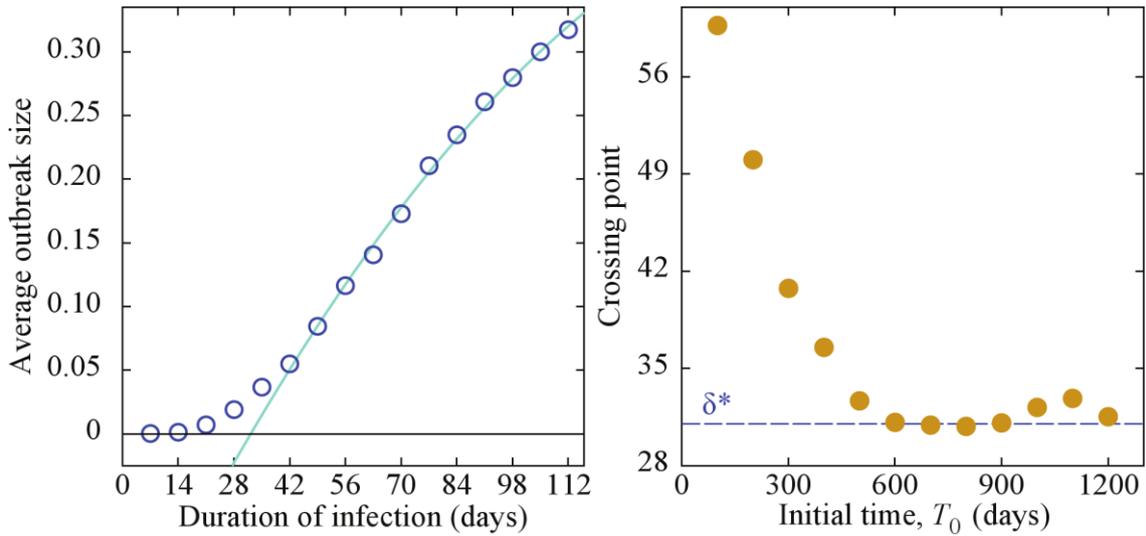

Figure 5. (*a*) Average outbreak size as a function of the duration of the infective stage. The line is a second-order polynomial fit to the data in the interval 42 days $\leq T \leq$ 112 days. (*b*) shows the time of such crossing point as a function of the beginning of the sampling window.

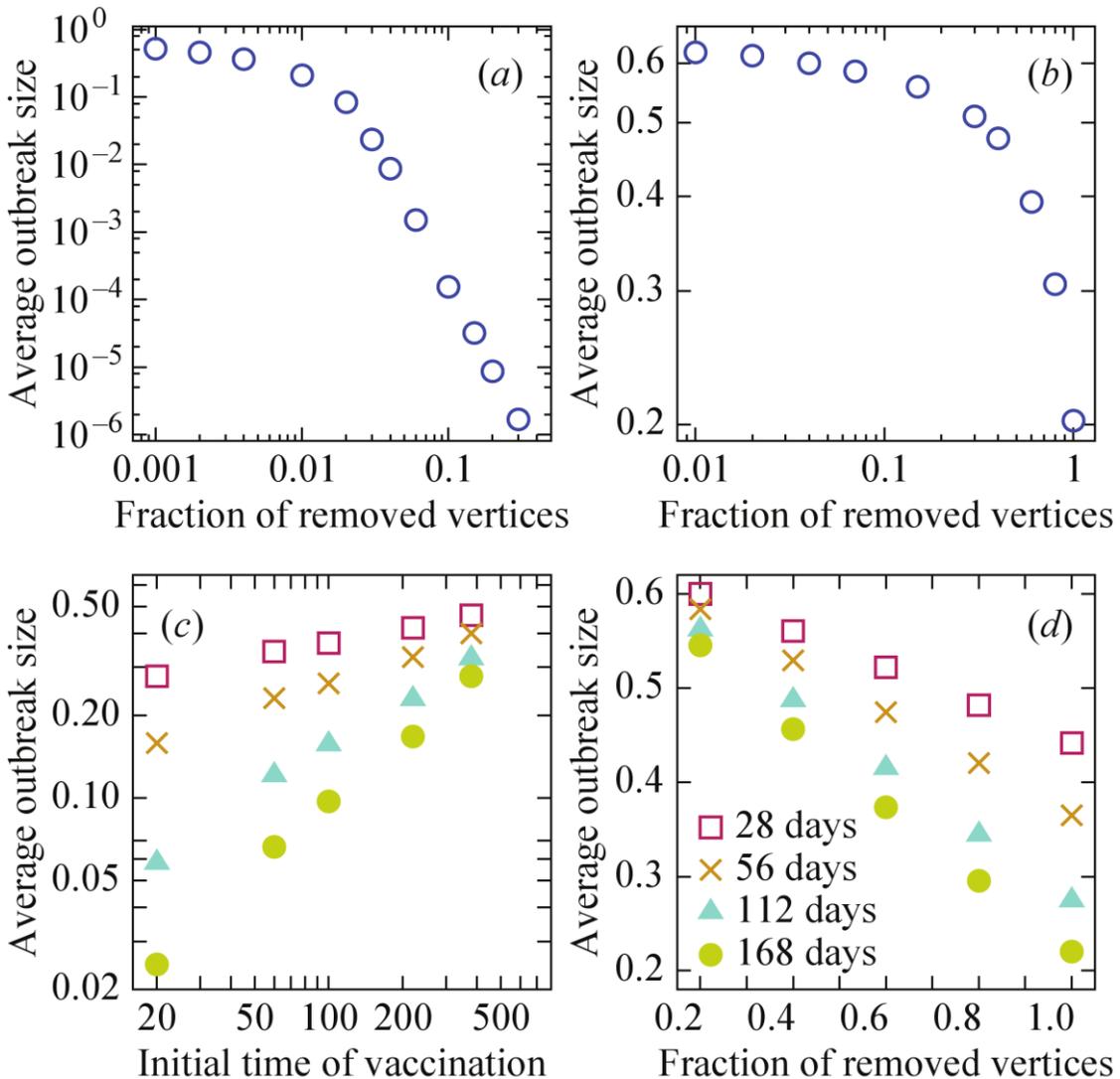

Figure 6. The average fraction of infected vertices after vaccination of (*a*) the most connected individuals; (*b*) the most travelled individuals; (*c*) all individuals during a fixed time interval $[\tau, \tau + \Delta\tau]$ (where $\tau$ is the time since the first infection); (*d*) a fraction of the individuals during an interval ($\Delta\tau = 300$ days). The disease simulations are by the SI model with $\rho = 1$.

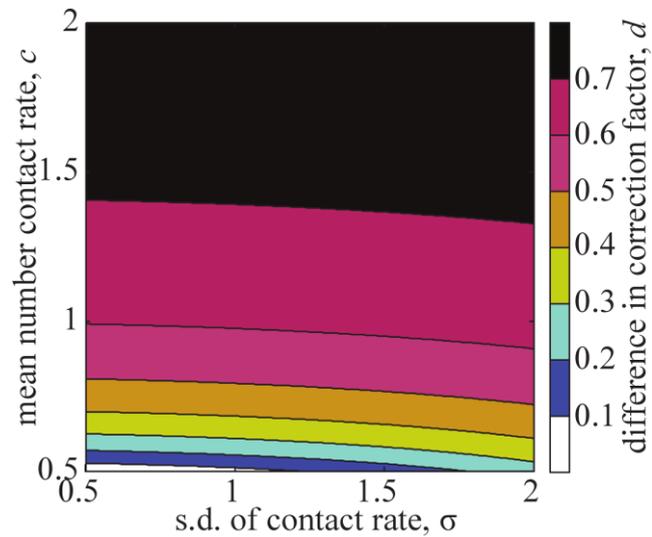

Figure 7. The difference $d$ in the correction factor to $R_0$ with and without our empirical network as a function of the average contact rate $c$ and the standard deviation of the contact rate $\sigma$ of the external contact patterns.